\documentclass[graybox]{svmult}

\usepackage{helvet}         
\usepackage{courier}        
\usepackage{type1cm}        
\usepackage{makeidx}         
\usepackage{graphicx}        
\usepackage{multicol}        
\usepackage[bottom]{footmisc}

\usepackage{amsmath}
\usepackage{amssymb}
\usepackage[utf8]{inputenc}
\usepackage{bbold}
\usepackage{xcolor}
\usepackage[makeroom]{cancel}
\usepackage{mathtools}
\usepackage{bbm}

\begin{document}

\title*{Unifying Gravities with Internal Interactions \vspace{0.3cm}\break
\small \textit{Dedicated to the memory of the Academician Ivan Todorov, a great scientist whose academic achievements have left an indelible mark on the field of Theoretical and Mathematical Physics.}}
\titlerunning{Unifying Gravities with Internal Interactions}
\author{Stelios Stefas and George Zoupanos}
\institute{Stelios Stefas \at Physics Department, National Technical University of Athens, Zografou Campus, 157 80, Zografou, Greece, \email{dstefas@mail.ntua.gr}
\and George Zoupanos \at Max-Planck Institut f\'ur Physik, Boltzmannstr. 8, 85 748 Garching/Munich, Germany; \\ Universit\"at Hamburg, Luruper Chaussee 149, 22 761 Hamburg, Germany; \\ Deutsches Elektronen-Synchrotron DESY, Notkestra{\ss}e 85, 22 607, Hamburg, \mbox{Germany}, \email{george.zoupanos@cern.ch}}
\maketitle
\abstract{Reviving the old proposal of describing gravity as a gauge theory first we describe the construction of the Conformal and the Noncommutative (Fuzzy) Gravities in a gauge-theoretic manner. Then stressing the fact that the tangent group of a curved manifold and the manifold itself do not necessarily have the same dimensions, we show how the above Gravities can be unified with the Internal Interactions, the latter based on the GUT $SO(10)$.}

\section{Introduction}
\label{sec:1}

For more than a century, theoretical physics has pursued the unification of all fundamental interactions, employing diverse methods and approaches across different eras to advance this goal. Kaluza and Klein proposed a framework unifying gravity and electromagnetism -- the two well-established interactions at that time -- by extending spacetime to five dimensions \cite{Kaluza:1921,Klein:1926}, i.e. they used the notion of extra dimensions as a unification tool. The use of higher-dimensional theories has gained renewed interest since the realisation that non-Abelian gauge theories could naturally emerge from similar settings \cite{Cho:1975sf, CHO1987358, Kerner:1968} and could be used in a unified description of gravity with the Standard Model (SM) of Elementary Particle Physics. More specifically, it was found that if the space-time manifold takes the form of a direct product $M_D = M_4 \times B$, where $B$ is a compact Riemannian manifold possessing a non-abelian isometry group $S$, following the dimensional reduction to four dimensions, one obtains gravity coupled to a Yang-Mills theory based on $S$ as the gauge group, along with scalar fields. Clearly, the most attractive feature of this scheme was the geometrical unification of gravity with the rest of the interactions, as well as the fact that it provided a natural explanation of gauge symmetries. Unfortunately, this minimal setup faced serious problems, such as the lack of a viable classical ground state with the assumed simple direct product structure and, more crucially for low-energy physics, it could not lead to chiral fermions in four dimensions after dimensional reduction \cite{Witten:1983}. A remarkable improvement appeared when Yang–Mills fields were introduced in the original action at the cost of abandoning the purely geometric unification, though maintaining the notion of extra dimensions. The latter option gave rise to another, although less ambitious, scheme the Coset Space Dimensional Reduction (CSDR) \cite{forgacs, KAPETANAKIS19924, Kubyshin:1989vd, MANTON1981502}, introduced by Forgacs and Manton (F-M), which could naturally lead to chiral fermions in four dimensions. Inspired by the work of F-M, Scherk and Schwarz (S-S) developed a very similar scheme with the reduction done on the group manifolds \cite{SCHERK197961}, which despite the fact that it cannot accommodate chiral fermions, it has been the basis of numerous developments in string model building. A clear lesson from the above studies is that in higher-dimensional Grand Unified Theory (GUT) frameworks that include both Yang–Mills fields and fermions \cite{FRITZSCH1975193, Georgi:1974sy}, the appearance of chiral fermions in four dimensions requires the total spacetime dimension to be of the form $4n+2$ \cite{CHAPLINE1982461}.

It should be noted that Superstring Theories (SSTs) (see e.g., refs. \cite{Green2012-ul, Lust:1989tj, polchinski_1998}), which were developed a bit later, again with the aim of unifying all fundamental interactions, dominated the investigation of higher-dimensional unification for decades. Moreover, they offered a consistent framework in higher dimensions, with the heterotic string theory \cite{GROSS1985253} -- formulated in ten dimensions -- standing out as a particularly attractive case. This theory naturally accommodates Grand Unified Theory (GUT) gauge groups such as $E_8 \times E_8$, whose dimensional reduction can, in principle, reproduce the SM. However, it should also be noted that experimental confirmation of all higher-dimensional frameworks is still lacking, despite the recent, very promising in this regard, developments in the CSDR framework \cite{Chatzistavrakidis:2009mh, Irges:2011de, Manolakos:2020cco, Manousselis_2004, Patellis:2024dfl}.

It should be noted that, prior to the revival of interest in higher-dimensional frameworks, another significant development emerged directly in four dimensions. This approach paved the way for a natural connection between gravity and gauge theories, thus pointing towards the unification of all interactions. In particular, the Standard Model of Particle Physics is based on gauge theories, while gravity has long been known as capable of being formulated as a gauge theory \cite{Ivanov:1980tw, Ivanov:1981wn, kibble1961, Kibble:1985sn, macdowell, Matsumoto, Sciama, stellewest, Umezawa, utiyama}. Interest in this perspective was renewed with developments in supergravity \cite{freedman_vanproeyen_2012, Ortín_2015}, which also rely on gauge principles, and has more recently been extended to Noncommutative (NC) gravity \cite{castellani, Chatzistavrakidis_2018, manolakosphd, Manolakos:2022universe, Manolakos:2023hif, Manolakos_paper1, Manolakos_paper2, roumelioti2407}.

Concerning the formulation of gravity as a gauge theory, Weyl \cite{weyl1929, weyl} took the first step by relating electromagnetism to local phase transformations of the electron field and introducing the vierbein formalism, which became crucial in gauge formulations of gravity. Utiyama \cite{utiyama} made the next important step by showing that gravity could be treated as a gauge theory of the Lorentz group $SO(1,3)$, although his introduction of the vierbein was somewhat ad hoc. This `weakness' was addressed by Kibble \cite{kibble1961} and Sciama \cite{Sciama}, who proposed gauging the full Poincar\'e group. Subsequent work by Stelle and West \cite{Kibble:1985sn, stellewest} led to more refined constructions based on the de Sitter $SO(1,4)$ or anti-de Sitter $SO(2,3)$ groups, which are isomorphic to the Poincar\'e, also incorporating spontaneous symmetry breaking (SSB) to recover Lorentz invariance. The conformal group $SO(2,4)$ was likewise central to the development of Conformal (CG) and Weyl (WG) gravities \cite{KAKU1977304, Roumelioti:2024lvn}, Fuzzy Gravity (FG) \cite{Chatzistavrakidis_2018, manolakosphd, Manolakos:2022universe, Manolakos:2023hif, Manolakos_paper1, Manolakos_paper2, roumelioti2407}, and their supersymmetric extensions in $N=1$ supergravity \cite{freedman_vanproeyen_2012, KAKU1977304}.

A more direct and ambitious proposal, aiming to unify all interactions within a gauge-theoretic framework, has been revived recently \cite{Chamseddine2010, Chamseddine2016, Konitopoulos:2023wst, Krasnov:2017epi, Manolakos:2023hif, Nesti_2010, Nesti_2008, Patellis:2025qbl, Patellis:2024znm, Roumelioti:2025cxi, Roumelioti:dubna, Roumelioti:2024lvn, noncomtomos}, inspired by earlier attempts \cite{Percacci:1984ai, Percacci_1991, Weinberg:1984ke}. It is based on the observation that the dimensions of the tangent group of a curved spacetime do not necessarily coincide with the dimensions of the manifold itself. This possibility suggests employing higher-dimensional tangent groups on four-dimensional spacetime, thereby facilitating a unified gauge-theoretic treatment of gravity and internal interactions. In this way, methods originally developed for higher-dimensional theories, such as the CSDR \cite{CHAPLINE1982461, Chatzistavrakidis:2009mh, forgacs, Irges:2011de, KAPETANAKIS19924, Kubyshin:1989vd, LUST1985309, Manolakos:2020cco, Manousselis_2004, MANTON1981502, Patellis:2024dfl, SCHERK197961}, can be adapted to this four-dimensional formulation. Nevertheless, challenges such as implementing simultaneous Weyl and Majorana conditions to obtain realistic chiral spectra also reappear in this setting \cite{CHAPLINE1982461, KAPETANAKIS19924}. More recently, a unified gauge framework has been constructed that brings together conformal gravity and internal interactions \cite{Konitopoulos:2023wst, Manolakos:2023hif, Patellis:2025qbl, Patellis:2024znm, Roumelioti:2025cxi, Roumelioti:dubna, Roumelioti:2024lvn}, and has been further extended to similarly unify noncommutative (fuzzy) gravity \cite{roumelioti2407}.

\section{Gauge-Theoretic Formulation of Conformal Gravity}
\label{sec:2}

To begin with, we recall that Einstein Gravity (EG) can be successfully formulated as a gauge theory of the Poincar\'e group \cite{kibble1961}. Greater conceptual clarity and elegance, however, arise from gauging either the de Sitter (dS) group $SO(1,4)$ or the anti--de Sitter (AdS) group $SO(2,3)$, both of which possess ten generators, the same number as the Poincar\'e group. In these cases, spontaneous symmetry breaking (SSB) to the Lorentz group $SO(1,3)$ is achieved by introducing an appropriate scalar field \cite{Kibble:1985sn, manolakosphd, Roumelioti:2024lvn, stellewest}. Moreover, the Poincar\'e, dS, and AdS groups are all subgroups of the conformal group $SO(2,4)$, which has fifteen generators. In \cite{Kaku:1978nz}, the gauge-theoretic formulation of gravity was extended to this full conformal group, giving rise to Conformal Gravity (CG). In that original construction, the breaking of CG to EG or to Weyl's scale-invariant gravity was implemented by imposing algebraic constraints on the gauge fields. In contrast, \cite{Roumelioti:2024lvn} achieved this breaking dynamically for the first time, by introducing a scalar field into the action and employing the Lagrange multiplier method.

The gauge theory of CG is based on $SO(2,4)$, which is isomorphic to $SU(4)$ and also to $SO(6)$ (in the following we work in Euclidean signature for convenience). There are two distinct paths one can follow in order to spontaneously break $SO(2,4)$ down to $SO(1,3)$.

\paragraph{Path I: Breaking via Scalars in the Vector Representation}\hfill \break
The first path, leading to EG, is to introduce a scalar field in the vector representation $\mathbf{6}$ of $SO(6)$. This scalar acquires a vacuum expectation value (vev) along its $\langle \mathbf{1} \rangle$ component, as seen from the branching of the $\mathbf{6}$ under the maximal subgroup $SO(5)$:
\begin{equation}
\label{SO6toSO5}
\begin{aligned}
SO(6) &\supset SO(5),\\
\mathbf{6} &= \mathbf{1} + \mathbf{5}\, .
\end{aligned}
\end{equation}
The unbroken $SO(5)$, which is isomorphic to $SO(2,3)$, may then break further to $SO(1,3)$ when a scalar in the $\mathbf{5}$ obtains a vev along its $\langle \mathbf{1} \rangle$ component under
\begin{equation}
\label{SO5toSU2SU2}
\begin{aligned}
SO(5) &\supset SU(2) \times SU(2),\\
\mathbf{5} &= (\mathbf{1},\mathbf{1}) + (\mathbf{2},\mathbf{2})\, ,
\end{aligned}
\end{equation}
where $SU(2)\times SU(2)$ is isomorphic to both $SO(4)$ and the Lorentz algebra $SO(1,3)$. Consequently, two scalar fields in the $\mathbf{6}$ of $SO(2,4)$ are required to achieve the complete breaking $SO(2,4) \longrightarrow SO(1,3)$ as detailed in \cite{Roumelioti:2024lvn}.

\paragraph{Path II: Direct Breaking via the Antisymmetric Representation}\hfill \break
As an alternative approach to the one mentioned above, one may break $SO(2,4)$ directly to $SO(1,3)$ in a single step by introducing a scalar field in the second-rank antisymmetric representation $\mathbf{15}$ of $SO(6)\cong SO(2,4)$. Depending on the choice of vacuum, this procedure yields either EG or WG, as will be clarified below.

The conformal algebra comprises fifteen generators, which in four-dimensional notation can be expressed as
\[
\{M_{ab},\; P_a,\; K_a,\; D\},
\]
corresponding to Lorentz transformations, translations, special conformal transformations, and dilatations, respectively. An $SO(2,4)$-valued gauge connection $A_\mu$ can be expanded on those generators as
\begin{equation}
A_\mu 
= \frac{1}{2}\omega_\mu{}^{ab} M_{ab} 
  + e_\mu{}^a P_a 
  + b_\mu{}^a K_a 
  + \tilde{a}_\mu D ,
\end{equation}
where $e_\mu{}^a$ is the vierbein, $\omega_\mu{}^{ab}$ the spin connection, $b_\mu{}^a$ the special conformal gauge field, and $\tilde{a}_\mu$ the dilatation gauge field. The corresponding field strength takes the form
\begin{equation}
\label{fieldstrengthconformal}
F_{\mu\nu}
= \frac{1}{2}R_{\mu\nu}{}^{ab} M_{ab}
 + \tilde{R}_{\mu\nu}{}^{a} P_a
 + R_{\mu\nu}{}^{a} K_a
 + R_{\mu\nu} D ,
\end{equation}
whose components' (including the usual 4D curvature and torsion tensors) explicit expressions can be found in \cite{Roumelioti:2024lvn}.

A parity conserving action, quadratic in the curvature, is then introduced:
\begin{equation}
S_{SO(2,4)} 
= a_{CG} \int d^4x \left[
\operatorname{tr}\,\epsilon^{\mu\nu\rho\sigma} \, m \phi \, F_{\mu\nu}F_{\rho\sigma}
+ \left(\phi^2 - m^{-2}\mathbb{1}_4\right)
\right],
\end{equation}
where $\phi$ is a scalar in the antisymmetric representation $\mathbf{15}$, $m$ a dimensionfull parameter and the trace is defined as $\mathrm{tr} \to \epsilon_{abcd}[\text{Generators}]^{abcd}$. Moreover, since $\phi$ is an algebra element, it, too, can be expanded as
\begin{equation}
\phi = \phi^{ab} M_{ab} + \tilde{\phi}^{a} P_a + \phi^{a} K_a + \tilde{\phi} D .
\end{equation}

Following \cite{Li:1973mq}, we work in a gauge where $\phi$ is diagonal,
\[
\phi = \mathrm{diag}(1,1,-1,-1),
\]
and is purely in the direction of the dilatation generator $D$:
\begin{equation}
\phi = \tilde{\phi}D \xrightarrow{\phi^2=m^{-2}\mathbb{1}_4} \phi = -2m^{-1}D .
\end{equation}
In this gauge, an SSB occurs and the action becomes
\begin{equation}
S = -2 a_{CG} \int d^4x\, \operatorname{tr}\,\epsilon^{\mu\nu\rho\sigma} F_{\mu\nu}F_{\rho\sigma}D ,
\end{equation}
where the gauge fields are rescaled as \(e \to m e\), \(b \to m b\), and \(\tilde{a} \to m\tilde{a}\). Using the expansion of \(F_{\mu\nu}\) and the conformal algebra, following some straightforward calculations one arrives at the Lorentz-invariant action \cite{Roumelioti:2024lvn}
\begin{equation}
\label{SO13action}
S_{SO(1,3)}
= \frac{a_{CG}}{4}\int d^4x\, 
\epsilon^{\mu\nu\rho\sigma}\epsilon_{abcd}\,
R_{\mu\nu}{}^{ab}R_{\rho\sigma}{}^{cd}.
\end{equation}

A key feature of this broken action is that $\tilde{a}_\mu$ does not appear in it. Thus we may set $\tilde{a}_\mu=0$, which simplifies field strengths related to the $P$ and $K$ generators:
\begin{equation}
\begin{aligned}
\tilde{R}_{\mu\nu}{}^a 
&= m\, T_{\mu\nu}^{(0)a}(e) - 2m^2 \tilde{a}_{[\mu} e_{\nu]}{}^a 
\;\longrightarrow\;
m\, T_{\mu\nu}^{(0)a}(e),\\[4pt]
R_{\mu\nu}{}^a 
&= m\, T_{\mu\nu}^{(0)a}(b) + 2m^2 \tilde{a}_{[\mu} b_{\nu]}{}^a 
\;\longrightarrow\;
m\, T_{\mu\nu}^{(0)a}(b),
\end{aligned}
\end{equation}
where $T_{\mu\nu}^{(0)a}$ denotes the torsion tensor in the Poincar\`e case. Since these terms are absent from the action too, we may also set them equal to zero
\[
\tilde{R}_{\mu\nu}{}^{a} = 0, 
\qquad
R_{\mu\nu}{}^{a} = 0,
\]
thus yielding a torsion-free theory. Likewise, the absence of the dilatation component curvature $R_{\mu\nu}$ from the action, allows us to set $R_{\mu\nu}=0$, which implies the constraint
\begin{equation}
\label{e-b-relation}
e_\mu{}^a b_{\nu a} - e_\nu{}^a b_{\mu a} = 0 .
\end{equation}
This relation motivates considering solutions in which $e_\mu{}^a$ and $b_\mu{}^a$ are algebraically related. Two important cases are the following.

\paragraph{Case A: $\boxed{b_\mu{}^{a} = a\, e_\mu{}^{a}}\rightarrow$ Einstein Gravity}\hfill \break  
\hfill \break  
This solution was proposed in \cite{Chamseddine:2002fd}, and by substituting it in \eqref{SO13action}, after some calculations one gets:
\begin{equation}
\begin{aligned}
S_{SO(1,3)}
= \frac{a_{CG}}{4}\int d^4x\, \epsilon^{\mu\nu\rho\sigma}\epsilon_{abcd}
\Big[
& R_{\mu\nu}^{(0)ab} R_{\rho\sigma}^{(0)cd}
- 16 m^2 a\, R_{\mu\nu}^{(0)ab} e_\rho{}^c e_\sigma{}^d \\
& + 64 m^4 a^2\, e_\mu{}^a e_\nu{}^b e_\rho{}^c e_\sigma{}^d
\Big],
\end{aligned}
\end{equation}
where the first term is a topological Gauss-Bonnet invariant (which does not affect field equations), the second term is the Einstein-Hilbert (Palatini) action equivalent in the vierbein formalism, and the third term plays the role of a cosmological constant term. In case $a < 0$, the above theory describes General Relativity in an AdS background spacetime.

\paragraph{Case B: $\boxed{b_\mu{}^{a} 
= -\frac{1}{4}\left(R_\mu{}^{a}-\frac{1}{6}R\, e_\mu{}^{a}\right)} \rightarrow$ Weyl Gravity} \hfill\break 
\hfill \break  
This solution is taken in \cite{Kaku:1978nz} and \cite{freedman_vanproeyen_2012}, and its substitution in \eqref{SO13action} yields, after some more calculations, an action made of the product of two Weyl conformal tensors $C_{\mu\nu}{}^{ab}$:
\begin{equation}
\label{Weyl1}
S = \frac{a_{CG}}{4} \int d^4x\, \epsilon^{\mu\nu\rho\sigma}\epsilon_{abcd}\, C_{\mu\nu}{}^{ab} C_{\rho\sigma}{}^{cd}\ .
\end{equation}
The above corresponds to the equally known form of scale-invariant Weyl action in four dimensions:
\begin{equation}
\label{Weyl2}
S_W =2a_{CG}\int \mathrm{d}^4 x\left(R_{\mu \nu} R^{\nu \mu}-\frac{1}{3} R^2\right).
\end{equation}
The Weyl action of WG in the forms given in eqs. \eqref{Weyl1} and \eqref{Weyl2}, being scale invariant, naturally does not contain a cosmological constant. WG is an attractive possibility for describing gravity at high scales (for some recent developments see \cite{Anastasiou:2016jix, Condeescu:2023izl, ghilencea2023, Hell:2023rbf, Maldacena:2011mk, mannheim}), as CG does. However, in the case that WG is obtained after the SSB of the CG, as is described above, a question remains on how one can obtain Einstein gravity from the SSB of WG. This can be done by introducing a scalar in the 2nd rank anti-symmetric tensor of $SU(4)\cong SO(2,4)$, $\mathbf{6}$, which after SSB leads to the E-H action \cite{Patellis:2025qbl, Patellis:2025syq}.

It should also be noted, as it was discussed explicitly in \cite{Patellis:2025qbl} that the vanishing of the two torsions $\tilde{R}_{\mu\nu}{}^{a}$ and $R_{\mu\nu}{}^{a}$ as well as the curvature tensor, $F_{\mu \nu}$ which is satisfied on-shell guarantee the equivalence of the diffeomorphisms and gauge transformations.

\section{Gauge-Theoretic Formulation of Noncommutative (Fuzzy) Gravity}
\label{sec:3}

Here we will briefly review the essential ingredients required for formulating gauge theories on noncommutative spaces, as these structures play a central role in what follows. Within the framework of noncommutative geometry, gauge fields appear quite naturally and are closely tied to the concept of a \emph{covariant coordinate} \cite{Madore:2000en}, which acts as the analogue of the covariant derivative in the commutative setting.

Let us consider a field $\phi(X_a)$ defined on a fuzzy space, depending on the noncommuting coordinates $X_a$. The field transforms in some representation of a gauge group $G$, and the infinitesimal gauge variation with parameter $\lambda(X_a)$ takes the form
\begin{equation}
    \delta \phi(X) = \lambda(X)\,\phi(X).
\end{equation}
When the gauge parameter $\lambda(X)$ is an ordinary function of the coordinates, the transformation is Abelian and the gauge group is $G = U(1)$. If instead $\lambda(X)$ is a $P \times P$ matrix, the transformation corresponds to a non-Abelian gauge symmetry with $G = U(P)$, the group of Hermitian $P \times P$ matrices. Importantly, the coordinates themselves are taken to be invariant under gauge transformations, i.e.\ $\delta X_a = 0$.

Applying a gauge transformation to the product of a coordinate and the field yields
\begin{equation}
    \delta(X_a \phi) = X_a \lambda(X)\, \phi,
\end{equation}
which is not covariant, since in general
\begin{equation}
    X_a \lambda(X)\phi \neq \lambda(X) X_a \phi.
\end{equation}
Motivated by the role of the covariant derivative in ordinary gauge theory, we introduce in the noncommutative context the \emph{covariant coordinate} $\mathcal{X}_a$, defined by the transformation rule
\begin{equation}
\label{3.5p}
    \delta(\mathcal{X}_a \phi) = \lambda\, \mathcal{X}_a \phi.
\end{equation}
This requirement is satisfied provided that
\begin{equation}
\label{3.6p}
    \delta \mathcal{X}_a = [\lambda, \mathcal{X}_a].
\end{equation}
The covariant coordinate is then written as
\begin{equation}
    \mathcal{X}_a \equiv X_a + A_a,
\end{equation}
allowing one to interpret $A_a$ as the gauge connection. Combining \eqref{3.5p} and \eqref{3.6p}, the gauge transformation of $A_a$ follows immediately:
\begin{equation}
    \delta A_a = -[X_a , \lambda] + [\lambda , A_a].
\end{equation}
This provides an a posteriori justification for viewing $A_a$ as the gauge field (for further details, see \cite{Aschieri:2004vh}).

This formalism will serve as the foundation for constructing a gravity model viewed as a gauge theory on a fuzzy covariant space in the following sections.

\subsection{The Noncommutative Background Space}\

Before constructing the gauge-theoretic formulation of Fuzzy Gravity, we must first specify the noncommutative background space on which the theory is defined. Following Snyder’s pioneering idea \cite{Snyder:1946qz} and its subsequent extensions \cite{Heckman_2015, Manolakos:2022universe, Manolakos_paper1, Manolakos_paper2, yang1947}, we adopt the viewpoint that spacetime itself is realized as a noncommutative manifold whose coordinate operators arise from the Lie algebra of the group $SO(1,5)$.

The generators $J_{mn}$ of $SO(1,5)$, with indices $m,n,r,s = 0,\dots,5$, satisfy the standard algebra
\begin{equation}
\left[J_{mn}, J_{rs}\right]
    = i \left( 
        \eta_{mr} J_{ns} + \eta_{ns} J_{mr}
        - \eta_{nr} J_{ms} - \eta_{ms} J_{nr}
    \right),
\end{equation}
where the metric is taken to be $\eta_{mn} = \mathrm{diag}(-1,1,1,1,1,1)$.

To uncover the 4D geometric interpretation, we decompose the group as
\[
SO(1,5) \supset SO(1,4) \supset SO(1,3).
\]
The resulting commutation relations are
\begin{equation}
\begin{gathered}
\left[J_{ij},J_{kl}\right]
    = i \left(
        \eta_{ik}J_{jl} + \eta_{jl}J_{ik}
        - \eta_{jk}J_{il} - \eta_{il}J_{jk}
    \right), \qquad
\left[J_{ij}, J_{k5}\right]
    = i \left(\eta_{ik}J_{j5} - \eta_{jk}J_{i5}\right), \\[4pt]
\left[J_{i5}, J_{j5}\right] = i J_{ij}, \qquad
\left[J_{ij}, J_{k4}\right]
    = i \left(\eta_{ik}J_{j4} - \eta_{jk}J_{i4}\right), \qquad
\left[J_{i4}, J_{j4}\right] = i J_{ij}, \\[4pt]
\left[J_{i4}, J_{j5}\right] = i \eta_{ij} J_{45}, \qquad
\left[J_{ij}, J_{45}\right] = 0, \qquad
\left[J_{i4}, J_{45}\right] = -i J_{i5}, \qquad
\left[J_{i5}, J_{45}\right] = i J_{i4}.
\end{gathered}
\end{equation}

These generators can be reinterpreted as physical quantities by identifying different generators of the algebra with the noncommutativity tensor, the spacetime coordinates, and the momenta:
\begin{equation}
\Theta_{ij} = \hbar J_{ij}, \qquad
X_i = \lambda J_{i5}, \qquad
P_i = \frac{\hbar}{\lambda} J_{i4}, \qquad
h = J_{45},
\end{equation}
where $\lambda$ is a length scale. Substituting these definitions back into the algebra yields the operator relations
\begin{equation}
\begin{gathered}
[\Theta_{ij}, \Theta_{kl}]
    = i\hbar \left(
        \eta_{ik}\Theta_{jl} + \eta_{jl}\Theta_{ik}
        - \eta_{jk}\Theta_{il} - \eta_{il}\Theta_{jk}
    \right), \\[4pt]
[\Theta_{ij}, X_k] = i\hbar (\eta_{ik} X_j - \eta_{jk} X_i),
\qquad
[\Theta_{ij}, P_k] = i\hbar (\eta_{ik} P_j - \eta_{jk} P_i), \\[4pt]
[X_i, X_j] = \frac{i\lambda^2}{\hbar} \Theta_{ij}, \qquad
[P_i, P_j] = \frac{i\hbar}{\lambda^2} \Theta_{ij}, \qquad
[X_i, P_j] = i \hbar \eta_{ij} h, \\[4pt]
[\Theta_{ij}, h] = 0, \qquad
[X_i, h] = \frac{i\lambda^2}{\hbar} P_i, \qquad
[P_i, h] = -\frac{i\hbar}{\lambda^2} X_i.
\end{gathered}
\end{equation}

This algebra encodes several crucial physical features. First, spacetime coordinates are manifestly noncommutative:
\[
[X_i, X_j] \propto \Theta_{ij},
\]
and the same holds for momenta. Both spacetime and momentum space therefore possess a discrete, quantized structure. Second, the mixed commutator $[X_i, P_j]$ reproduces a generalized Heisenberg relation.

In this way, the Snyder--Yang algebra emerges as the natural starting point for a consistent noncommutative background suitable for defining a gauge theory of Fuzzy Gravity.

\subsection{Gauge Theory of Fuzzy Gravity}\ 

Having established the noncommutative background space, we now proceed to formulate a gauge theory of gravity defined on it. The first step will be to identify an appropriate gauge group. A natural starting point is the isometry group of $dS_4$, namely $SO(1,4)$, which governs the symmetries of the corresponding commutative gravitational theory. However, in noncommutative gauge theories the appearance of anticommutators between gauge generators is unavoidable. Since the anticommutators of $SO(1,4)$ generators do not, in general, close within the original algebra, one must work in a representation for which closure under both commutators and anticommutators is ensured. 

Following the procedure presented in \cite{Manolakos_paper1, Manolakos_paper2}, the generators' representation is fixed and the gauge group must be enlarged from $SO(1,4)$ to $SO(2,4)\times U(1)$, the minimal group in which the generators as well as both the commutators and anticommutators close. With the gauge group established, we can now construct the noncommutative gauge theory of Fuzzy Gravity on the covariant Snyder--Yang background.

\vspace{0.3cm}

We begin by introducing the covariant coordinate
\begin{equation}\label{CovariantCoordinate}
\mathcal{X}_\mu = X_\mu \otimes \mathbb{1}_4 + A_\mu(X),
\end{equation}
where $A_\mu$ is the noncommutative gauge connection. Expanding $A_\mu$ on the generators of $SO(2,4)\times U(1)$ gives
\begin{equation}\label{GaugeConnectionFuzzy}
A_\mu
    = a_\mu \otimes \mathbb{1}_4
    + \omega_\mu{}^{ab}\otimes M_{ab}
    + e_\mu{}^a \otimes P_a
    + b_\mu{}^a \otimes K_a
    + \tilde{a}_\mu \otimes D.
\end{equation}
Substituting this expansion into \eqref{CovariantCoordinate}, the covariant coordinate takes the explicit form
\begin{equation}
\mathcal{X}_\mu
    = (X_\mu + a_\mu)\otimes \mathbb{1}_4
    + \omega_\mu{}^{ab}\otimes M_{ab}
    + e_\mu{}^a \otimes P_a
    + b_\mu{}^a \otimes K_a
    + \tilde{a}_\mu \otimes D.
\end{equation}

The covariant noncommutative field strength tensor is then defined as \cite{Madore_1992, Manolakos_paper1}
\begin{equation}
\hat{F}_{\mu\nu} \equiv [\mathcal{X}_\mu, \mathcal{X}_\nu] - \kappa^2 \hat{\Theta}_{\mu\nu}.
\end{equation}
Above we have defined: 
\[
\hat{\Theta}_{\mu\nu} \equiv \Theta_{\mu\nu} + \mathcal{B}_{\mu\nu},
\]
where $\mathcal{B}_{\mu\nu}$ is a 2-form field which takes care of the covariance of $\Theta_{\mu\nu}$, and $\kappa^2$ turns out to be equal to $\frac{i\lambda^2}{\hbar}$ . As an element of the gauge algebra, the field strength admits the decomposition
\begin{equation}
\hat{F}_{\mu\nu}= R_{\mu\nu}\otimes \mathbb{1}_4 + \frac{1}{2} R_{\mu\nu}{}^{ab}\otimes M_{ab} + \tilde{R}_{\mu\nu}{}^a \otimes P_a + R_{\mu\nu}{}^a \otimes K_a + \tilde{R}_{\mu\nu}\otimes D.
\end{equation}

To obtain a physically relevant gravitational theory, a spontaneous symmetry breaking (SSB) is performed, in close analogy with the conformal gravity construction. One introduces a scalar field $\Phi(X)$ transforming in the second-rank antisymmetric representation of $SO(6)\cong SO(2,4)$, and additionally assigns it a $U(1)$ charge so that the relevant part of the extended symmetry is fully broken. Fixing the scalar in an appropriate gauge reduces the local symmetry from $SO(2,4)\times U(1)$ down to the Lorentz group $SO(1,3)$ \cite{Manolakos_paper1, Manolakos_paper2, Roumelioti:2024lvn}. 

The resulting action for Fuzzy Gravity is then found to be
\begin{equation}
\mathcal{S} = \operatorname{Trtr} \Big[ \lambda\, \Phi(X)\, \varepsilon^{\mu\nu\rho\sigma} \hat{F}_{\mu\nu}\hat{F}_{\rho\sigma} + \eta\Big( \Phi(X)^2 - \lambda^{-2}\mathbb{1}_N\otimes\mathbb{1}_4 \Big) \Big],
\end{equation}
where $\eta$ is a Lagrange multiplier and $\lambda$ is a dimensionful parameter. After the symmetry breaking, the residual gauge symmetry is $SO(1,3)$. As demonstrated in \cite{Manolakos_paper2}, the commutative limit of this action reproduces the Palatini action, and hence ordinary Einstein gravity with a cosmological constant is retrieved.

\section{SO(2,16) Unification of Gravities and Internal Interactions}
\label{sec:4}

In \cite{Roumelioti:2024lvn}, it was proposed that CG can be unified with internal interactions within a framework that naturally leads to an $SO(10)$ Grand Unified Theory (GUT), by using $SO(2,16)$ as the single unification gauge group. This choice for the unification gauge group is motivated from the facts that:
\begin{itemize}
    \item It should be possible to reach both the $SO(2,4)$ and $SO(10)$ gauge groups through SSBs, starting from the initial unification gauge group and
    \item In order to have a chiral theory, we need a group of the form $SO(4n+2)$.
\end{itemize}
Given the above requirements, it becomes evident that the smallest unification group which satisfies them is the $SO(2,16)$. As highlighted in the Introduction, the key idea relies on the fact that the dimension of the tangent group need not match with that of the underlying manifold \cite{Chamseddine2010, Chamseddine2016, Konitopoulos:2023wst, Krasnov:2017epi, Nesti_2010, Nesti_2008, Percacci:1984ai, Percacci_1991, roumelioti2407, noncomtomos, Weinberg:1984ke}.  

In what follows, for reasons of simplicity, we work with the Euclidean signature (the implications of the non-compact case is discussed in detail in \cite{Roumelioti:2024lvn}). Starting from $SO(18)\cong SO(2,16)$ with fermions in its spinor rep, $\mathbf{256}$, the SSB leads to its maximal subgroup $SO(6)\times SO(12)$ \cite{Roumelioti:2024lvn}. The relevant branching rules are:
\begin{equation}\label{so18}
\begin{aligned}
SO(18) & \supset SO(6) \times SO(12) \\
\mathbf{256} & = (\mathbf{4}, \overline{\mathbf{32}}) + (\overline{\mathbf{4}}, \mathbf{32}) 
  && \text{(spinor)} \\
{\mathbf{153}} & =(\mathbf{15}, \mathbf{1}) + (\mathbf{6}, \mathbf{12}) + (\mathbf{1}, \mathbf{66}) & & \text {(adjoint)} \\
\mathbf{170} & = (\mathbf{1}, \mathbf{1}) + (\mathbf{6}, \mathbf{12}) + (\mathbf{20}', \mathbf{1}) + (\mathbf{1}, \mathbf{77}) 
  && \text{(2nd rank symmetric)}
\end{aligned}
\end{equation}

The breaking of $SO(18)$ to $SO(6) \times SO(12)$ is triggered by assigning a vev to the $(\mathbf{1},\mathbf{1})$ component of a scalar in the $\mathbf{170}$ rep with fermions in the $\mathbf{256}$ spinor rep of $SO(18)$.

In order to further break $SO(12)$ down to $SO(10) \times U(1)$ or $SO(10) \times U(1)_{\text{global}}$, we can employ scalar fields from the $\mathbf{66}$ representation (contained within the adjoint $\mathbf{153}$ of $SO(18)$) or the $\mathbf{77}$ representation (contained within the 2nd rank symmetric tensor representation $\mathbf{170}$ of $SO(18)$), respectively, given the following branching rules:
\begin{equation}
\begin{aligned}
SO(12) & \supset SO(10) \times \left[U(1)\right]\\
\mathbf{66} & =(\mathbf{1})(0)+( \mathbf{10})(2)+( \mathbf{10})(-2)+( \mathbf{45})(0) \\
\mathbf{77} & =(\mathbf{1})(4)+( \mathbf{1})(0)+( \mathbf{1})(-4)+( \mathbf{10})(2)+( \mathbf{10})(-2)+( \mathbf{54})(0) \\
\end{aligned}
\end{equation}
where the $\left[{U}(1)\right]$ above is there to take into account that the $U(1)$ either remains as a gauge symmetry, or it is broken leaving a $U(1)$ as a residual global symmetry. Given the above branching rules, a VEV to the $\langle(\mathbf{1})(0)\rangle$ component of the $\mathbf{66}$ representation leads to the gauge group $SO(10) \times U(1)$ after the SSB, while a VEV to the $\langle(\mathbf{1})(4)\rangle$ component of the $\mathbf{77}$ representation results in $SO(10) \times U(1)_{\text{global}}$.

Similarly, we can further break $SU(4)\cong SO(6)$ down to $SO(4) \cong SU(2) \times SU(2)$ in two stages, following the same procedure described in Path I in Sec. \ref{sec:2} \cite{Slansky:1981yr}: \\ 
As an initial step, according to the branching rules \eqref{SO6toSO5}, by assigning a VEV to the $\langle\mathbf{1}\rangle$ component of a scalar in the $\mathbf{6}$ representation of $SU(4)$, the latter breaks down to $SO(5)$. Then, according to the branching rules \eqref{SO5toSU2SU2}, by giving a VEV to the $(\mathbf{1},\mathbf{1})$ component of a scalar in the $\mathbf{5}$ representation of $SO(5)$, we finally obtain the Lorentz group $SU(2) \times SU(2) \cong SO(4) \cong SO(1,3)$. 

Additionally, it is notable that in this scenario, the $\mathbf{4}$ representation decomposes under $SU(2) \times SU(2) \cong SO(1,3)$ into the appropriate representations to describe two Weyl spinors.

One can also follow an alternative route to break $SU(4)$ to $SU(2) \times SU(2)$, just like in the CG case. Specifically, in order to break the $SU(4)$ gauge group to $SU(2) \times SU(2)$, we can use scalars in the adjoint $\mathbf{15}$ representation of $SU(4)$, which is contained in the adjoint $\mathbf{153}$ representation of $SO(18)$. In this case, we have:
\begin{equation}
\begin{aligned}
SU(4) \supset & SU(2) \times SU(2) \times U(1)\\
\textbf{4} = & (\textbf{2},\textbf{1})(1)+(\textbf{1},\textbf{2})(-1)\\
\textbf{15}  = & (\textbf{1},\textbf{1})(0)+(\textbf{2},\textbf{2})(2)+(\textbf{2},\textbf{2})(-2)\\
&+(\textbf{3},\textbf{1})(0)+(\textbf{1},\textbf{3})(0),
\end{aligned}
\end{equation}
from where, by assigning a VEV to the $(\mathbf{1},\mathbf{1})$ direction of the adjoint  representation $\mathbf{15}$, we obtain the known result \cite{Li:1973mq} that $SU(4)$ spontaneously breaks to $SU(2) \times SU(2) \times U(1)$. The method for eliminating the corresponding $U(1)$ gauge boson and retaining only $SU(2) \times SU(2)$ is the same as in the CG case. Again, note that the $\mathbf{4}$ representation decomposes into the appropriate representations of $SU(2) \times SU(2) \cong SO(1,3)$ suitable for describing two Weyl spinors.

Having established the analysis of various symmetry breakings using branching rules under maximal subgroups, starting from the group $SO(18)$, one can correspondingly consider instead the isomorphic algebras of the various groups. Specifically, instead of $SO(18)$, one can consider the isomorphic algebra of the non-compact groups $SO(2,16) \cong SO(18)$, and similarly $SO(2,4) \cong SO(6) \cong SU(4)$.

Consequently, after all the breakings, we obtain:
\begin{equation}
\begin{gathered}
    SU(2)\times SU(2) \times SO(10) \times [U(1)]\\
    \{(\textbf{2},\textbf{1})+(\textbf{1},\textbf{2})\}\{\textbf{16}(-1)+\overline{\textbf{16}}(1)\}+\{(\textbf{2},\textbf{1})+(\textbf{1},\textbf{2})\}\{\overline{\textbf{16}}(1)+\textbf{16}(-1)\}\\
    =2\times\textbf{16}_L(-1)+2\times\overline{\textbf{16}}_L(1)+2\times\textbf{16}_R(-1)+2\times\overline{\textbf{16}}_R(1), 
\end{gathered}
\end{equation}
from where, given that $\overline{\textbf{16}}_R(1)=\textbf{16}_L(-1)$ and $\overline{\textbf{16}}_L(1)=\textbf{16}_R(-1)$, and by choosing to keep only the $\textbf{-1}$ eigenvalue of $\gamma^5$, we obtain
\begin{equation}
    4\times \textbf{16}_L(-1)\, .
\end{equation}
Therefore, this construction yields a natural prediction of four fermion families, arising from the underlying group-theoretic structure. The flavour separation is left as an open problem for future work.

For the Fuzzy Gravity case, in ref \cite{roumelioti2407} it is noted that unifying FG with internal interactions requires fermions to:
\begin{itemize}
    \item be chiral to survive at low energies and not acquire masses at the Planck scale, and 
    \item appear in matrix reps consistent with the matrix model construction of FG.
\end{itemize}

The suggested resolution can be outlined as follows: rather than placing fermions in fundamental or tensor representations of $SU(N)$ gauge groups, one may consider fermions in bi-fundamental representations of product gauge groups. This strategy can similarly be applied to $SO(N)$ gauge theories. Constructions of this type have appeared previously in various contexts and with different  goals \cite{Ibanez:1998xn, Irges:2011de, Leontaris:2005ax, Ma:2004mi, Manolakos:2020cco, Patellis:2024dfl}.

In the current current case, this is achieved by starting with the $SO(6)\times SO(12)$ gauge theory and fermions in $(\mathbf{4}, \overline{\mathbf{32}})+(\overline{\mathbf{4}}, \mathbf{32})$, thus satisfying both criteria. Additionally, the gauge-theoretic formulation of gravity in FG requires gauging $SO(2,4) \times U(1) \cong SO(6)\times U(1)$, leading to a low-energy structure closely analogous to the CG case.

\section{Conclusions}
\label{sec:5}

In \cite{Roumelioti:2024lvn}, a potentially realistic framework was developed in which gravity and internal interactions in four dimensions are unified by gauging an enlarged tangent Lorentz group. This approach relies on the key observation that the dimension of the tangent group can exceed that of the underlying manifold. By constructing CG as a gauge theory of $SO(2,4)$ and implementing spontaneous symmetry breaking, both Einstein Gravity and Weyl Gravity emerge as possible low-energy limits. 

Extending this framework to include internal interactions via $SO(10)$ GUTs was achieved using the higher-dimensional tangent group $SO(2,16)$, with fermions subject to the Weyl condition. A parallel construction for Fuzzy Gravity \cite{roumelioti2407} starts from $SO(2,4)\times SO(12)$ with fermions in $(\mathbf{4}, \overline{\mathbf{32}})+(\overline{\mathbf{4}}, \mathbf{32})$, leading to a unified, gauge-theoretic description of fuzzy gravity and internal interactions.

The low energy limit of the above construction has been studied in \cite{Patellis-Z-24}, by the employment of a 1-loop analysis. Four channels of breaking $SO(10)$ down to the SM have been explored, providing estimates for all the breaking scales from the Planck scale down to the EG scale as well as the possible signals of the related cosmic strings in the gravitational waves \cite{Patellis:2025qbl}.

This article is dedicated to the Academician Ivan Todorov, whose memory will remain with us forever. We express our deepest condolences to his wife Boryana, his family, colleagues, and students at the Institute for Nuclear Research and Nuclear Energy in Sofia, as well as to all those worldwide who continue to carry forward his spirit.

\end{document}